\documentclass[preprint,aps,groupedaddress,floatfix,onecolumn,superscriptaddress,showpacs]{revtex4-1}
\usepackage{amsmath,amssymb,epsfig,bm,pifont}
\usepackage[linkcolor=black,citecolor=black,urlcolor=blue,colorlinks=true,linktocpage=true]{hyperref}
\usepackage{graphicx}
\usepackage{epstopdf}
\usepackage{bm}
\usepackage{tabularx}

\newcommand{\beq}{\begin{equation}}
\newcommand{\eeq}{\end{equation}}
\newcommand{\bea}{\begin{eqnarray}}
\newcommand{\eea}{\end{eqnarray}}

\newcommand{\veps}{\varepsilon}

\begin{document}

\title{Hard-Wall and Non-Uniform Lattice Monte Carlo Approaches to One-Dimensional  Fermi Gases in a Harmonic Trap}

\author{Casey E. Berger}
\email{cberger3@live.unc.edu}
\affiliation{Department of Physics and Astronomy, University of North Carolina, Chapel Hill, NC, 27599, USA}

\author{Joaqu\'{\i}n E. Drut}
\email{drut@email.unc.edu}
\affiliation{Department of Physics and Astronomy, University of North Carolina, Chapel Hill, NC, 27599, USA}

\author{William J. Porter}
\email{wjporter@live.unc.edu}
\affiliation{Department of Physics and Astronomy, University of North Carolina, Chapel Hill, NC, 27599, USA}

\begin{abstract}
We present in detail two variants of the lattice Monte Carlo method aimed at tackling systems in external trapping potentials: a uniform-lattice approach with 
hard-wall boundary conditions, and a non-uniform Gauss-Hermite lattice approach. Using those two methods, we compute the ground-state energy and spatial 
density profile for systems of $N=4 - 8$ harmonically trapped fermions in one dimension. From the favorable comparison of both energies and density profiles 
(particularly in regions of low density), we conclude that the trapping potential is properly resolved by the hard-wall basis. Our work paves the way to 
higher dimensions and finite temperature analyses, as calculations with the hard-wall basis can be accelerated via fast Fourier transforms; the cost of 
unaccelerated methods is otherwise prohibitive due to the unfavorable scaling with system size.\end{abstract}

\date{\today}
\maketitle

\section{Introduction}

\begin{figure}[h!]
\includegraphics[width=1.0\columnwidth]{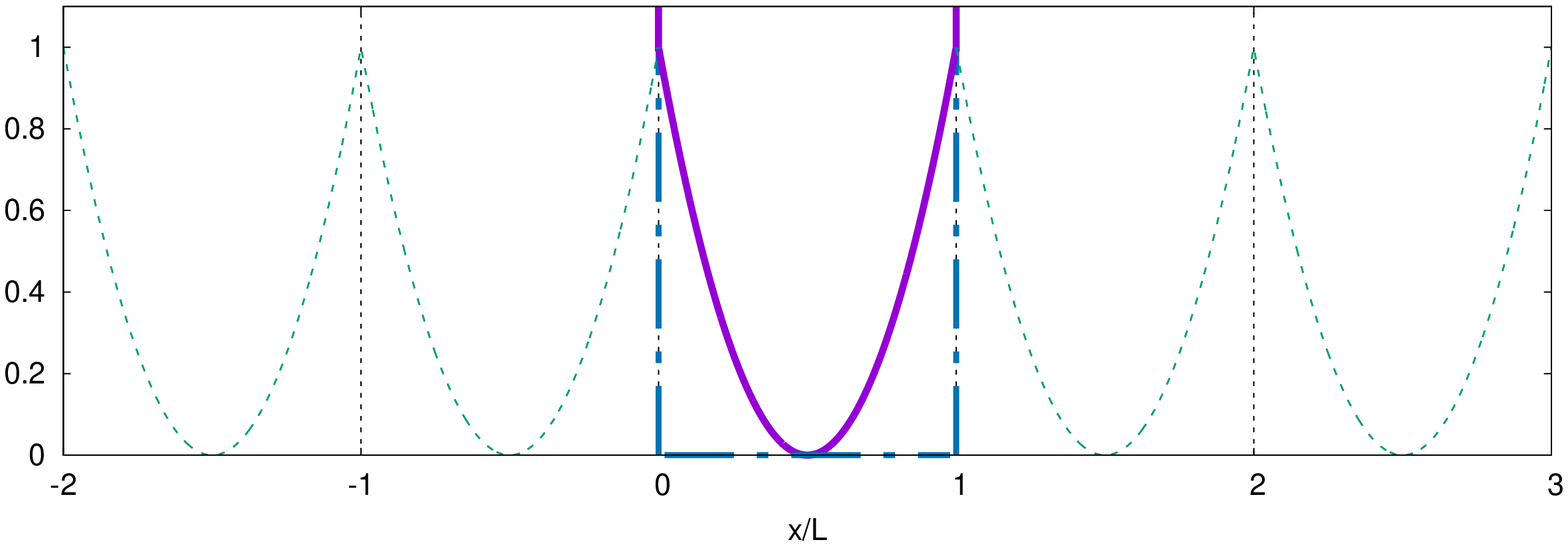}
\caption{\label{Fig:ExtPot} Schematic view of a harmonic potential in a one-dimensional box with periodic boundary conditions (dashed lines) 
as contrasted with the same potential but with hard-wall boundaries (region  $0 < x/L < 1$; solid line). The hard-wall potential itself is shown
with dashed-dotted lines.
}
\end{figure}

The quantum Monte Carlo method has been around for nearly as long as modern computers 
(see e.g. Ref.~\cite{MCReviews1, MCReviews2, MCReviews3} for reviews). 
By far, most calculations that use that method, from 
condensed-matter and ultracold-atom systems to quantum chromodynamics, are performed in uniform lattices with periodic boundary conditions. 
This approach makes sense in most of those cases, as the aim is to describe nearly uniform systems, which are such that periodic boundary conditions minimize finite-size effects. However, this is not always a good approximation in the case 
of ultracold atoms, where the optical trapping potential plays a central role in experiments and dictates the many-body properties of the 
system~\cite{RevExp,RevTheory1,RevTheory2}.
It is therefore essential to include a harmonic trap in realistic calculations. As a result of this inclusion, translation invariance is broken and 
plane waves on a uniform periodic lattice are no longer the natural 
basis of the system. Indeed, in the presence of a trap, momentum ceases to be a good quantum number. Moreover, the boundary conditions of the true harmonic oscillator are not at all periodic; in fact, implementing periodicity would result in undesirable copies of the harmonic potential across the boundaries 
(see Figure~\ref{Fig:ExtPot}).

To resolve the above issues, we attempted in Ref.~\cite{GH-BergerAndersonDrut} to use the natural coordinate-space lattice of the harmonic oscillator, namely the Gauss-Hermite points and weights of gaussian quadrature methods. The dual basis in this case is of course that of harmonic-oscillator wavefunctions. Although such a non-uniform lattice is physically and mathematically attractive, it is not efficient from the computational standpoint: the scaling with the system size is prohibitive in more than a single spatial dimension, and there appears to be no simple way to accelerate those calculations as Fourier transforms do for uniform lattices (there are, however, possible routes via non-uniform fast Fourier transforms~\cite{NFFT}).

In this work, we carry out a test of a methodological compromise between the choices mentioned above: we return to the uniform-lattice basis, but implement it with hard-wall boundary conditions (i.e. an infinite square-well potential). The latter prevent the appearance of spurious copies of the harmonic potential across the boundary, while at the same time allow for Fourier acceleration (with a small, sub-leading cost of linearly combining the results of Fourier transforms). As a proof of principle, we compute properties of trapped 1D fermions, namely the ground-state energy and density profiles, and compare with calculations in the non-uniform basis. Although much is known about fermions in 
1D in uniform space~\cite{Review1D}, most previous works have combined the classic Bethe-ansatz solution with the local-density approximation 
in order to treat trapped systems~\cite{HomPlusLDA1,HomPlusLDA2,HomPlusLDA3,HomPlusLDA4,HomPlusLDA5,HomPlusLDA6,BAplusLDA}. (See however Refs.~\cite{ExactDiag1,ExactDiag2,ExactDiag3} for exact-diagonalization approaches.) Our goal here, in contrast,
is to design a more general Monte Carlo method to account for the trapping potential in an {\it ab initio} fashion, which we could apply in higher dimensions.


\section{General Overview of the Methods}

Our calculations explore the properties of a system of $N$ nonrelativistic, equal-mass, $\mbox{SU}(2)$ fermions in one spatial dimension under the 
influence of an external harmonic potential $V^{}_0(x) = (1/2)m \omega^2 x^2$ with mass $m$ and trap frequency $\omega$.  We take the particles 
to have dispersion $\veps(p) = p^2/(2m)$ and to interact via an attractive, pair-wise, short-range potential.  Specifically, we study the Hamiltonian 
$\hat{H}$ written
\beq
\label{Eq:Hamiltonian}
\hat{H} = \hat{T} + \hat{V}_{0} + \hat{V}
\eeq
where $\hat{T}$ is a one-body kinetic energy operator, $\hat{V}_{0}^{}$ is coupling to our static background, and $\hat{V}$ is the interparticle potential.  
Throughout, we work in units where $k_{B}^{} = \hbar = m = 1$.

In both approaches, we place our system on a discrete spacetime of dimensionless size $N_{x}^{} \times N_{\tau}^{}$.  Where we employ simple harmonic oscillator basis 
states, the system has no well-defined physical volume, and as a result, the length and momentum scales are set by the frequency $\omega$.  Further, for these 
calculations we choose a nonuniform lattice spacing to be described in detail below.  By contrast, studies performed in the square-well basis are endowed with a natural 
volume, and in this instance, we work with a uniform spatial lattice of size $L = N_{x}^{}\ell$, taking $\ell = 1$ throughout to fix the relevant scales. Although the details of 
the spatial discretization differ between the two approaches, the temporal lattice is uniform and of dimensionful extent $\beta = N_{\tau}^{}\tau$.

We begin, in each method, with a trial state $|\Omega\rangle$ and obtain the many-body, ground-state expectation value of an operator $\hat{O}$ via large-imaginary-time projection.  Specifically, for eigenfunctions $|E_n\rangle$ of $\hat{H}$ and assuming $\langle E_0|\Omega \rangle$ is nonvanishing, it follows from completeness that
\beq
O_\beta \equiv \frac{\langle \Omega(\beta/2) |\hat{O}| \Omega(\beta/2) \rangle}{\langle \Omega(\beta/2) | \Omega(\beta/2) \rangle }\xrightarrow[]{\beta\to\infty} \langle E_0^{}|\hat{O}|E_0^{} \rangle,
\eeq
where
\beq
| \Omega(\tau) \rangle = \hat{U}(\tau,0)|\Omega\rangle,
\eeq
and where we have defined the imaginary-time evolution operator
\beq
\hat{U}(\tau_b^{},\tau_a^{}) = e^{-(\tau_b^{} - \tau_a^{})\hat{H}}.
\eeq

Our representation of each operator comprising $\hat{H}$ given in Equation~(\ref{Eq:Hamiltonian}) is method-specific, and the details of the lattice Monte Carlo (MC) 
technique are in each case intuitively tied to our choice of basis.  In both of the methods discussed below, we partition the Hamiltonian into two noncommuting 
operators $\hat{H} = \hat{H}_0 + \Delta \hat{H}$, approximating the typical MC projectors via a symmetric Suzuki-Trotter (ST) decomposition in order to isolate a 
single-particle piece $\hat{H}_0^{}$ whose exponential we can explicitly diagonalize. In the non-uniform basis method, $\hat{H}_0 = \hat T + \hat V_0$ (diagonal 
in harmonic-oscillator space), whereas in the uniform hard-wall basis we take $\hat{H}_0 = \hat T$ (diagonal in momentum space). Generically, we approximate 
each factor comprising the evolution operator as
\beq
\exp\left[\tau\left(\hat{H}_0 + \Delta \hat{H}\right)\right] = \exp\left(-\frac{\tau}{2}\hat{H}_0\right)\exp\left(-\tau\Delta\hat{H}\right)\exp\left(-\frac{\tau}{2}\hat{H}_0\right) + \mathcal O(\tau^3)
\eeq
for small $\tau$.

In both cases, the factors involving $\hat{H}_0$ are implemented in diagonal form, but in order to tackle the central factor, we implement a Hubbard-Stratonovich (HS) transformation~\cite{HS1,HS2} to decouple the central two-body observable.  Schematically, we write
\beq
\label{Eq:HubbardStrat}
\exp\left(-\tau\Delta\hat{H}\right) = \int \mathcal D \sigma(x,\tau_i) \exp\left(-\Delta\hat{H}_{\sigma(i)}\right)
\eeq
for each point on the imaginary-time lattice where we have introduced a spatially fluctuating HS auxiliary field $\sigma(x,\tau_i)$ and a collection of one-body operators $\Delta\hat{H}_{\sigma(i)}$.  Gathering the path integrals, we may write the composite evolution operator as an integral over a (now space-time varying) field as
\beq
\label{Eq:UDecomp1}
\hat{U}(\beta,0) = \int \mathcal D \sigma(x,\tau)\;\hat{M}_{\sigma}+\mathcal O(\tau^2)
\eeq
where
\beq
\label{Eq:UDecomp2}
\hat{M}_{\sigma} = \prod_{i=N_\tau}^{1}\exp\left(-\frac{\tau}{2}\hat{H}_0\right)\exp\left(-\Delta\hat{H}_{\sigma(i)}\right)\exp\left(-\frac{\tau}{2}\hat{H}_0\right).
\eeq
Application of the matrices $\hat{M}_\sigma$ for a each configuration of the HS field constitutes a sizable component of the calculation, and by repeatedly switching 
between two separate bases, the action of each factor is computed using its diagonal representation.  
The above sequence of transformations (Trotter-Suzuki, Hubbard-Stratonovich) leads to the path-integral representation of $O^{}_\beta$, which we evaluate using 
Metropolis-based Monte Carlo methods, in particular hybrid Monte Carlo~\cite{HMC1,HMC2}.
Further details are presented below, and for a more complete 
discussion see Refs.~\cite{MCReviews1,MCReviews2,MCReviews3}.


\section{Technical Aspects of the Methods}

\subsection{Non-uniform lattice method}

Through its connection to gaussian quadrature methods~\cite{StoerBulirsch, NR}, this partition of the Hamiltonian, or equivalently the choice of which basis functions to use in the above-mentioned diagonization, provides a natural lattice geometry.  Specifically, the need to resolve the chosen basis states, as well as projections onto them, with high precision motivates a prudent choice of not only the orbitals themselves but also of the integration method.  Expanding a generic trial state demonstrates immediately that in order to guarantee faithful resolution of this state in terms of single-particle orbitals, it is sufficient to ensure that the orthonormality of the basis is preserved.  We perform our calculations in each case on the lattice corresponding to a quadrature appropriate to the basis at hand.  In this way, we maintain exactly the orthonormality of the single-particle wavefunctions and the fidelity of our computations expressed thereby.

Written entirely in position space, we have the kinetic energy operator
\beq
\hat{T} = \sum_{s = \uparrow,\downarrow} \int dx\;\hat{\psi}^{\dagger}_{s}(x)\;\veps\left(\frac{1}{i}\frac{\partial}{\partial x}\right)\;\hat{\psi}^{}_{s}(x)
\eeq
expressed via field operators $\hat{\psi}^{}_{s}(x)$ and $\hat{\psi}^{\dagger}_{s}(x)$ for a state specified by position and spin quantum numbers $(x,s)$, as well as the two-body contact interaction 
\beq
\hat{V} = -g\int dx \; \hat{n}^{}_{\uparrow}(x)\hat{n}^{}_{\downarrow}(x)
\eeq
 given in terms of the density operators $\hat{n}_{s}^{} = \hat{\psi}^{\dagger}_{s}\hat{\psi}^{}_{s}$ with nonnegative bare coupling $g$  and the static background potential
\beq
\hat{V_0} = \sum_{s = \uparrow,\downarrow} \int dx \; V^{}_0(x)\; \hat{n}_{s}^{}(x).
\eeq

As described in Ref.~\cite{GH-BergerAndersonDrut}, a convenient basis is the set of single particle orbitals $\alpha_k^{}(x)$ satisfying
\beq
\left(-\frac{\partial^2}{\partial \xi_{}^2} + \xi_{}^2\right)\alpha_k^{} = (2k+1)\alpha_{k}^{}
\eeq
for dimensionless variable $\xi = \sqrt{\omega} x$ and nonnegative integer $k$, corresponding to the harmonic oscillator (HO).  Expanding the field operators in terms of creation and annihilation operators $b_{k,s}^{\dagger}$ and $b_{k,s}^{}$ corresponding to HO quantum numbers $(k,s)$ as
\beq
\hat{\psi}^{(\dagger)}_{s}(x) = \sum_{k=0}^{\infty} \alpha_{k}^{}(x)\hat b_{k,s}^{(\dagger)},
\eeq
 we diagonalize the first two summands comprising $\hat{H}$ as
\beq
\label{Eq:SHODiag}
\hat{T} + \hat{V}_{0} = \sum_{s=\uparrow,\downarrow} \sum_{k=0}^{\infty} \omega \left(k+\frac{1}{2}\right)\;\hat b_{k,s}^{\dagger}\hat b_{k,s}^{}.
\eeq
Using this diagonal form, we perform the calculations using the nonlinear lattice (described below) by implementing the operator given in Equation~(\ref{Eq:SHODiag}), in HO space, the remaining contact interaction $\hat{V}$ in position space, and switching between throughout the application of Equations~(\ref{Eq:UDecomp1}),(\ref{Eq:UDecomp2}).  In order to efficiently represent the HO single-particle orbitals $\alpha_k^{}(x)$ in coordinate space, we place the system on a lattice corresponding to $N_x$ Gauss-Hermite (GH) integration points $x_{i}$ (with associated weights $w_i > 0$) in lieu of the more conventional uniform discretization associated with calculations performed using a basis of momentum eigenstates.

Indeed, a real-valued function $f(x)$ sampled on an $n$-site GH lattice may be numerically integrated (see Refs.~\cite{NR, StoerBulirsch}), often to exceptional accuracy, via
\beq
\label{Eq:GHInt}
\int_{-\infty}^{\infty}dx \; f(x)e^{-x^2} = \sum_{i = 1}^{n} w_i f(x_i) + \frac{n!\sqrt{\pi}}{2^n\;(2n)!} f^{(2n)}(\zeta)
\eeq
with weights
\beq
w_i = \frac{2^{n-1}n!\sqrt{\pi}}{n^2\;[H_{n-1}(x_i)]^2},
\eeq
for real $\zeta$, and where the values $x_i$ are determined by the roots of the $n$-th order Hermite polynomial $H_n(x)$.
These conditions are derived by requiring the sum in Equation~(\ref{Eq:GHInt}) to exactly reproduce the desired integral when the function 
$f$ is taken to be a polynomial of degree $\deg f < 2n$.  By choosing to represent our system in coordinate space on a $N_x$-site 
spatial lattice, we maintain exactly that the first $N_x$ HO orbitals form an orthonormal set.

\subsection{Uniform lattice with hard-wall boundary method}

Although the previous approach is attractive in its elegant preservation of system's underlying structure even after discretization, its scaling, particularly in comparison to conventional Fourier-accelerated MC approaches (see Ref.~\cite{FourierAcceleration1, FourierAcceleration2}), places discouraging limits on this method's applicability vis-\`a-vis higher dimensional systems.  In any dimension, the scaling is determined by the computational cost of matrix-vector operations, which naively scales quadratically in the lattice volume, that is $\mathcal{O}(V^2)$ for $V = N_x^d$. Accelerated calculations using a uniform lattice, however, achieve scaling as benign as $\mathcal{O}(V\ln V)$~\cite{GH-BergerAndersonDrut}.  Additionally, the Fourier-transform basis is naturally orthonormal on a uniform lattice making it all the more appealing.

Computational cleverness and simplicity aside, the basis functions associated with conventional uniform-lattice techniques exhibit boundary conditions that differ dramatically from those characterizing eigenstates of the system at hand.  For any finite system size, decomposition in periodic functions fails to capture the required asymptotic behavior, specifically that the density must be localized in space and must eventually vanish monotonically as the distance from the trap's center grows.  Although they do not exhibit the same type of decay and despite being compactly defined, the eigenfunctions corresponding to the infinite square well (SW), that is a system confined by a hard-wall (HW) trapping potential,  do vanish at the system's boundaries.  Further, even though the GH lattice is defined to represent functions defined on the entire real line, for any finite lattice size, it inevitably fails to capture effects coming from the long-distance tails where the discrete representation of the function is no 
longer supported.  Judicious use of this technique circumvents the problem almost entirely, as these omissions are minimal when the function of interest is localized near the origin.

In light of the above, we propose studying a harmonically trapped gas using a large uniform lattice, in the sense that $ L \sqrt{\omega}  \gg 1$, but rather than making use of the conventional plane-wave decomposition, we work in a basis of SW wave functions $\phi_n(x)$ for positive integers $n$, supported for $0 < x < L$, 
and defined by
\beq
\phi_{n}(x) = \sqrt{\frac{2}{L}}\sin \left(\frac{n\pi x}{L}\right).
\eeq
Expanding the field operators in terms of Fock-space operators which destroy (respectively, create) a SW state with quantum numbers $(n,s)$, denoted $a_{n,s}^{(\dagger)}$, as
\beq
\hat{\psi}^{(\dagger)}_{s}(x) = \sum_{n=1}^{\infty} \phi_{n}^{}(x)\hat a_{n,s}^{(\dagger)},
\eeq
we may diagonalize the kinetic energy operator alone to find
\beq
\hat{T} = \sum_{s = \uparrow,\downarrow} \sum_{n=1}^{\infty} \frac{p_{n}^{2}}{2m}\;\hat a_{n,s}^{\dagger}\hat a_{n,s}^{}
\eeq
where we have written the SW momenta as $p_{n} = \pi n/L$.  As is conventionally done, we apply the remaining operators, those derived from the interparticle interaction and from the presence of the background potential, in position space where they are diagonal after the HS transformation.
Since $\phi_{n}(x)$ is a linear combination of (conventional, complex-exponential) plane waves, it is straightforward to take advantage of fast Fourier 
transform algorithms to accelerate these HW calculations. 
Indeed, that linear combination relating $\phi_{n}$ to complex exponentials involves only $2^d$ terms, i.e. it is a sparse operation whose scaling is only linear with 
the lattice volume $V=N_x^d$. In particular, it scales more favorably than a general change of basis, the cost of which is $O(V^2)$.


\section{Results and Discussion}

\begin{figure}[t]
\includegraphics[width=1.0\columnwidth]{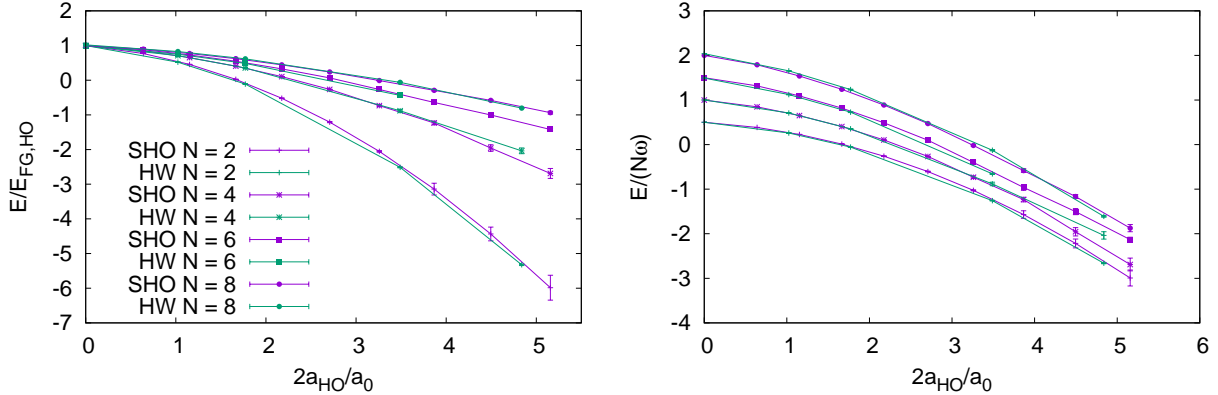}
\caption{\label{Fig:EnergyPlots} Energy per particle in units of the energy of the non-interacting case (left panel)
and in units of the harmonic-oscillator energy $\hbar \omega$ (right panel), for $N=2,4,6,8$ particles (bottom to top), 
as obtained with the harmonic-oscillator basis (SHO) and uniform hard-wall basis (HW).
}
\end{figure}

To tune the bare coupling in our lattice calculations, we first computed the ground-state energy of the two-body problem.
Doing so allowed us to read off the value of the continuum physical coupling (as given by the inverse scattering length $1/a_0^{}$),
as the exact solution of the two-body system in a harmonic trap can be obtained exactly and is well-known~\cite{BuschEtAl}.
Having fixed the target physics in that fashion, for both methods, we were able to meaningfully compare the results obtained with
each of them for higher particle number.
In Figure~\ref{Fig:EnergyPlots}, for instance, we show our results for the coupling tuning procedure ($N = 2$), which are exact by definition, 
along with results obtained for higher particle numbers ($N = 4, 6, 8$) for those couplings.

\begin{figure}[h!]
\includegraphics[width=1.0\columnwidth]{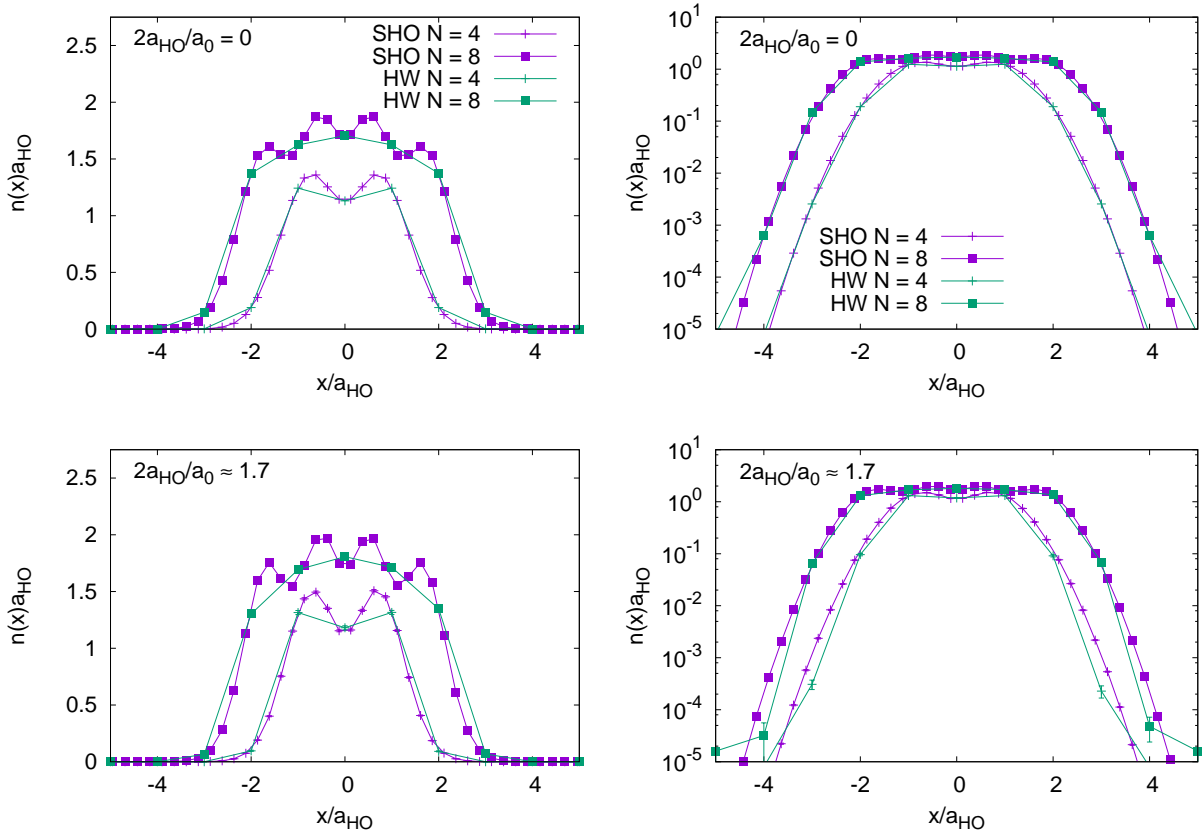}
\caption{\label{Fig:DensityPlots} Density profiles for noninteracting systems (top) and interacting (bottom).
The left panels display the data in linear $y$ scale, while the right panels show a $y$-log scale. 
In all cases we show results for $N=4$ and $8$ particles, and for the harmonic-oscillator basis (SHO) and 
uniform hard-wall basis (HW).
}
\end{figure}

Figure~\ref{Fig:DensityPlots} shows density profiles for noninteracting systems of $N=4$ and $8$ particles, along with
their counterparts for an interacting case at $2a^{}_\text{HO}/a^{}_0\simeq1.7$. The left panels in that figure show the
profiles in a linear scale, whereas the right panels show them in a $y$-log scale. In all cases we see that, whenever
the $x$ axis values coincide (or do so approximately) the results for density have the expected values, i.e. the two approaches 
agree quantitatively. The logarithmic plots also show excellent agreement; more precisely, we see that the long-distance
tails (in each direction) decay in a parabolic form, which indicates that the dependence in the linear scale is gaussian, 
as expected. Note, however, that at large enough distances that parabolic form is lost for the hard-wall data, which is not
surprising given the presence of the walls.

%


\section{Summary and Conclusions}

In this paper, we have presented two methods to address the problem of interacting fermions in harmonic traps:
a uniform-lattice method with hard-wall boundary conditions, and a non-uniform Gauss-Hermite lattice method (which
we had used in previous work). While the latter has many attractive features (it diagonalizes the 
noninteracting Hamiltonian exactly), it is not amenable to Fourier acceleration (or at least not easily), which
makes it practically unfeasible for higher dimensions (especially away from zero temperature).
The hard-wall method, on the other hand, shares some of the positive features and can be Fourier accelerated,
as we explained above.

To test the methods against each other, we compared here calculations for 1D attractively interacting fermions 
in a harmonic trap. Specifically, we computed the ground-state energy and density profiles of unpolarized systems 
of $N=4$ and $8$ particles.
Our results show that for both the ground-state energy and the density profiles, the methods agree 
satisfactorily. For the density profiles, in particular, we note that the expected gaussian decay is reproduced 
with the hard-wall basis over multiple orders of magnitude before breaking down at large distances due to the 
presence of the wall.
From our calculations we conclude that it is possible to obtain high-quality results using uniform bases with
hard-wall boundaries. 

Besides the above benefits, the hard-wall method has the advantage that it does not depend on the precise form
of the external potential. Indeed, it is easy to imagine that it would be a useful method for other trapping potentials that
are unbounded at infinity (e.g. linear or other). Moreover, the hard-wall potential is interesting per se, as experiments
with ultracold atoms can now mimic that kind of configuration as well (albeit with somewhat rounded corners at
the bottom of the trap, which could be introduced quite easily in our framework).


\acknowledgments{}

We acknowledge discussions with E. R. Anderson and with L. Rammelm\"uller.
This material is based upon work supported by the National Science Foundation under 
Grants 
No. PHY1306520 (Nuclear Theory program) 
and 
No. PHY1452635 (Computational Physics program).




\begin{thebibliography}{999} 

\bibitem{MCReviews1}
F. F. Assaad and H. G. Evertz,
Worldline and Determinantal Quantum Monte Carlo Methods for Spins, Phonons and Electrons, in 
{\it Computational Many-Particle Physics}, H. Fehske, R. Shnieider, and A. Weise Eds., Springer, Berlin (2008);

\bibitem{MCReviews2}
D. Lee, Phys. Rev. C {\bf 78}, 024001(2008);
D. Lee, Prog. Part. Nucl. Phys. {\bf 63}, 117 (2009).

\bibitem{MCReviews3}
J. E. Drut and A. N. Nicholson, J. Phys. G {\bf 40}, 043101 (2013).


\bibitem{RevExp}
\textit{Ultracold Fermi Gases}, 
Proceedings of the International School of Physics ``Enrico Fermi", Course CLXIV, 
Varenna, June 20 -- 30, 2006,
M.~Inguscio, W.~Ketterle, C.~Salomon (Eds.) (IOS Press, Amsterdam, 2008).


\bibitem{RevTheory1}
I. Bloch, J. Dalibard, W. Zwerger, Rev. Mod. Phys. {\bf 80}, 885 (2008).

\bibitem{RevTheory2}
S. Giorgini, L. P. Pitaevskii, S. Stringari, Rev. Mod. Phys. {\bf 80} 1215 (2008).

\bibitem{GH-BergerAndersonDrut}
C. E. Berger, E. R. Anderson, J. E. Drut, Phys. Rev. A {\bf 91}, 053618 (2015).

\bibitem{NFFT}
\url{www.nfft.org}

\bibitem{Review1D}
X-W. Guan, M. T. Batchelor, C. Lee, Rev. Mod. Phys. {\bf 85}, 1633 (2013).

\bibitem{HomPlusLDA1}
I. V. Tokatly, Phys. Rev. Lett. {\bf 91}, 090405 (2004).

\bibitem{HomPlusLDA2}
G. E. Astrakharchik, D. Blume, S. Giorgini, L. P. Pitaevskii, Phys. Rev. Lett. {\bf 93}, 050402 (2004).

\bibitem{HomPlusLDA3}
H. Hu, X.-J. Liu, P. D. Drummond, Phys. Rev. Lett. {\bf 98}, 070403 (2007).

\bibitem{HomPlusLDA4}
G. Orso, Phys. Rev. Lett. {\bf 98}, 070402 (2007).

\bibitem{HomPlusLDA5}
P. Kakashvili, C. J. Bolech, Phys. Rev. A {\bf 79}, 041603(R) (2009).

\bibitem{HomPlusLDA6}
J.-H. Hu, J.-J. Wang, G. Xianlong, M. Okumura, R. Igarashi, S. Yamada, M. Machida, Phys Rev. B {\bf 82}, 014202 (2010).

\bibitem{BAplusLDA}
S. Schenk, M. Dzierzawa, P. Schwab, U. Eckern, Phys. Rev. B {\bf 78}, 165102 (2008).

\bibitem{ExactDiag1}
P. D'Amico, M. Rontani, Phys. Rev. A {\bf 91}, 043610 (2015).

\bibitem{ExactDiag2}
T. Sowi\'nski, M. Gajda, K. Rz\c a\.zewski, EPL {\bf 109}, 26005 (2015).

\bibitem{ExactDiag3}
E. J. Lindgren, J. Rotureau, C. Forss\'en, A. G. Volosniev, N. T. Zinner, New J. Phys. {\bf 16}, 063003 (2014).

\bibitem{HS1}
R. L. Stratonovich, Sov. Phys. Dokl. {\bf 2}, 416 (1958).

\bibitem{HS2}
J. Hubbard, Phys. Rev. Lett. {\bf 3}, 77 (1959).

\bibitem{HMC1}
S. Duane, A. D. Kennedy, B. J. Pendleton, D. Roweth, Phys. Lett. B {\bf 195}, 216 (1987).

\bibitem{HMC2}
S. A. Gottlieb, W. Liu, D. Toussaint, R. L. Renken, Phys. Rev. D {\bf 35}, 2531 (1987).

\bibitem{NR}
Press, W. H. , {\it et al.}, {\it Numerical Recipes in FORTRAN},
(2$^\text{nd}$ Ed., Cambridge University Press, Cambridge, England, 1992). 

\bibitem{StoerBulirsch}
Stoer, J.; Bulirsch, R.
Introduction to Numerical Analysis. 
Springer-Verlag: New York, USA, 2002; pp. 171-180. 

\bibitem{FourierAcceleration1} 
G. Batrouni, A. Hansen, M. Nelkin, Phys. Rev. Lett. {\bf 57}, 1336 (1986).

\bibitem{FourierAcceleration2}
C. Davies, G. Batrouni, G. Katz, A. Kronfeld, P. Lepage, P. Rossi, B. Svetitsky, K. Wilson, J. Stat. Phys. {\bf 43}, 1073 (1986).







\bibitem{BuschEtAl}
T. Busch, B.-G. Englert, K. Rz\c a\.zewski, M. Wilkens, Foundations of Physics {\bf 28}, 549 (1998).

%
%

\end{thebibliography}
\end{document}